\documentclass[article,11pt,apl,amsmath,amssymb,showpacs]{revtex4-2}
\usepackage{epsf}      
\usepackage{graphicx}
\usepackage{color}
\usepackage{soul}
\usepackage{gensymb}
\usepackage{sidecap}
\usepackage{amsmath}
\usepackage{mathtools}
\usepackage[hidelinks,colorlinks=true,linkcolor=blue,citecolor=blue]{hyperref}

\begin{document}
\title{Electrochemical performance and diffusion kinetics of a NASICON type Na$_{3.3}$Mn$_{1.2}$Ti$_{0.75}$Mo$_{0.05}$(PO$_4$)$_3$/C cathode for low-cost sodium-ion batteries}
\author{Madhav Sharma}
\affiliation{Department of Physics, Indian Institute of Technology Delhi, Hauz Khas, New Delhi-110016, India}
\author{Rajendra S. Dhaka}
\email{rsdhaka@physics.iitd.ac.in}
\affiliation{Department of Physics, Indian Institute of Technology Delhi, Hauz Khas, New Delhi-110016, India}

\date{\today}
\begin{abstract}
We report the electrochemical performance and diffusion kinetics of a newly designed NASICON type Na$_{3.3}$Mn$_{1.2}$Ti$_{0.75}$Mo$_{0.05}$(PO$_4$)$_3$/C composite material as a cathode for cost-effective sodium-ion batteries. A novel strategy of small Mo doping successfully stabilizes the sample having high Mn content in single phase rhombohedral symmerty. The high-resolution microscopy analysis reveals nanocrystallites of around $\sim$18 nm, uniformly embedded within the semi-graphitic carbon matrix, which enhances the surface electronic conductivity and effectively shortens the sodium-ion diffusion path. More importantly, we demonstrate a stable electrochemical behavior, with enhanced  discharge capacity of 124 mAh/g at 0.1 C, having good reversibility and retaining 77\% of its capacity after 300 cycles, and 70\% even after 400 cycles at 2 C. The sodium-ion diffusion coefficients, estimated using both galvanostatic intermittent titration technique (GITT) and cyclic voltammetry are found to lie within the range of $10^{-9}$ to $10^{-11}$~cm$^2$/s.  Additionally, the bond-valence site energy mapping predicted a sodium-ion migration energy barrier of 0.76 eV. A detailed distribution of relaxation times (DRT) analysis is used to deconvolute the electrochemical impedance spectra into distinct processes based on their characteristic relaxation times. Notably, the solid-state diffusion of sodium ions within the bulk electrode, with a relaxation time of $\sim$50 s, shows a consistent trend with the diffusion coefficients obtained from GITT and Warburg-based evaluations across the state of charge.  
\end{abstract}
\maketitle

\section{\noindent~Introduction}

Sodium-ion batteries (SIBs) are quickly emerging as a promising alternative to lithium-ion based energy storage devices, thanks to the abundance, low cost, and wide availability of sodium \cite{Chen_ASS_18, Rudola_NE_23}. Their working principles closely resemble, making the transition to develop cost-effective SIBs both feasible and attractive, which offer their practical route toward large-scale stationary energy storage applications \cite{Delmas_AEM_18, Abraham_ACSEL_20}. However, there are still many challenges due to the larger size of Na$^+$ ions where the associated structural strain during cycling can hinder the long-term performance of SIBs. Therefore, the search for stable and high-energy cathode materials is vital \cite{Madhav_CCR}. Among the available options, the polyanionic structured materials stand out for their robust frameworks, thermal stability, and minimal volume changes during Na$^+$ insertion and extraction, making them strong contenders for next-generation SIB technology \cite{Sapra_WEE_21, Lu_AM_24}. In this line, the Na$_3$V$_2$(PO$_4$)$_3$ and Na$_3$V$_2$(PO$_4$)$_2$F$_3$ are well-known members of the sodium (Na) SuperIonic CONductor (NASICON) family, and have been widely explored as cathode materials for SIBs over the recent years \cite{Simran_NVPF_arxiv}. Their popularity stems from their stable voltage profile, decent capacity, and long cycle life \cite{Wang_JMCA_18}. However, despite these advantages, concerns over the high cost and environmental impact of vanadium have raised challenges similar to those faced for cobalt in LIBs and pose significant challenges for the widespread adoption of vanadium-based SIBs \cite{Soundharrajan_JMCA_22}. These concerns have driven the search for new cathode materials consisting of abundant Fe and Mn that offer promising solutions for sustainable, low-cost, and eco-friendly options and can eliminate the reliance on vanadium while maintaining performance and stability of SIBs \cite{Madhav_CCR, Xu_ACSCS_23}. 

In this context, utilizing the Mn-based polyanionic cathode materials can be a viable solution \cite{Khan_JES_25, Niu_CN_2023}; however, they often suffer from significant polarization owing to the gap between the charge/discharge curve and limited cycling stability \cite{Liu_ESM_23}. The cathode materials like NaMnPO$_4$, Na$_2$MnPO$_4$F, Na$_2$MnP$_2$O$_7$, Na$_3$Mn$_2$(P$_2$O$_7$)(PO$_4$), and Na$_4$Mn$_3$(PO$_4$)$_2$P$_2$O$_7$ are studied extensively \cite{Boyadzhieva_RSCA_20, Wu_JPS_18, Barpanda_JMCA_13, Li_JPS_22, Kim_EES_15}; but the unavailability of Mn$^{3+}$/Mn$^{4+}$ redox couple poses a concern on the further enhancement of their electrochemical performance. Moreover, the process of cycling in Mn-based cathodes led to substantial deterioration due to the presence of the active Jahn-Teller (J-T) distortion, which resulted in structural instabilities \cite{Zheng_AEM_24}. In the NASICON framework, Na$_x$Mn$_2$(PO$_4$)$_3$ has been reported to be stable thermodynamically \cite{Singh_JMCA_21}, but there are no experimental reports to the best of our knowledge. On the other hand, in the case of binary transition metal (TM) NASICONs, the Mn was found to be outstand with every other transition metal in terms of cost, redox voltage, and number of participating sodium ions in storage per TM benefitting from the availability of Mn$^{3+}$/Mn$^{4+}$ redox \cite{Zhou_NL_16, Gao_JACS_18, Wang_AEM_20, Wang_CEJ_21}. In addition, when half of the Mn sites are substituted with other appropriate TMs, the J-T distortion is diminished, leading to improved cycling stability \cite{Gao_JACS_18}. Interestingly, the Mn-based NASICON cathodes are reported to have reversible execution of Mn$^{3+}$/Mn$^{4+}$ just 0.5 V above the Mn$^{2+}$/Mn$^{3+}$ redox, providing high energy density \cite{Wang_AEM_20}. However, one of the two TM sites in the structure is restricted to have an oxidation state of 3+ or higher to stabilize Mn$^{2+}$. The Ti (4+) with its d$^0$ configuration shows high stability during Ti$^{3+}$/Ti$^{4+}$ redox, and enhanced sodium kinetics, can be an ideal choice \cite{GuO_EES_16}. Here, the Na$_3$MnTi(PO$_4$)$_3$ (NMTP) with decent sodium concentration is capable of providing two sodium ions during charging in its pristine state \cite{Gao_CM_16}. When lowering the cutoff voltage below 2.1 V, it is capable of inserting another sodium ion via Ti$^{3+}$/Ti$^{4+}$ redox \cite{Zhu_AEM_19}. In the potential window of 1.5-4.3 V, the NMTP can give three sodium ion reactions at an average voltage of 3.0 V. 

Notably, the NASICON structure with rhombohedral symmetry can theoretically have four sodium ions per formula unit; for example, the basic charge balance calculations suggests that cathode material with Na$_4$Mn$_{1.5}$Ti$_{0.5}$(PO$_4$)$_3$ configuration satisfy the selection rules \cite{Liu_JMCA_21, Snarskis_CM21}. However, it was  concluded that increasing Mn concentration more than 1.15 leads to unwanted impurity phases \cite{Liu_JMCA_21} and considered a tedious task as the bigger size of Mn$^{2+}$ is well known to bring along deformations in the structure, which eventually leads to the deviation from the expected crystal symmetry \cite{Wu_JMC_11}. Also, in the polyanionic class of electrode materials, the electronic transport follows the polaronic type, which causes low conductivity and hampers the efficiency of the electrochemical process \cite{Johannes_PRB_12, Luong_JSAMD_17}. Additionally, the hole polaron state of Mn$^{3+}$ is situated within the strong-bonding e$_g$ complex and supports a J-T distortion, which results in an overall stronger local deformation of the lattice, leading to a higher migration barrier for electronic conduction at higher Mn content \cite{Luong_JSAMD_17}. In addition to that, the intrinsic anti-site disorder (IASD) caused by the site hopping of Mn$^{2+}$ between the TM site and alkali metal (AM) site in the NASICON framework causes a significant voltage hysteresis during the charge/discharge and introduces a notable redox pair: 4.0 V during charge and 2.5 V during discharge, which hampers the energy efficiency \cite{Zhang_ACSEL_21}. It was found that the Na$^+$/Mn$^{2+}$ cation mixing can be significantly reduced by filling the vacant AM sites with excess sodium \cite{Zhang_ACSEL_21, Xu_AFM_23, Zhang_CEJ_23} as well as by a TM doping strategy to restore the voltage hysteresis \cite{Liu_NE_23, Zhang_AM_24}. 

Herein, we successfully elevate the Mn content and synthesized Na$_{3.3}$Mn$_{1.2}$Ti$_{0.75}$Mo$_{0.05}$(PO$_4$)$_3$/C, a mid-entropy cathode material using sol-gel method and opted both strategies of increased sodium content and Mo doping to stabilize high Mn content, strengthen electrochemical reversibility and facilitate stable multi electron reaction. A characteristic sloping voltage profile is observed, likely due to the high surface state density associated with the material’s nanocrystalline nature, as evidenced by high-resolution transmission electron microscopy (HR-TEM). The material exhibits promising electrochemical performance, with an initial discharge capacity of 124 mAh/g at 0.1 C, good rate capability up to 5 C, and 70\% capacity retention after 400 cycles at 2 C, highlighting its potential as a stable and efficient SIB cathode. To investigate sodium ion transport, the diffusion coefficient values are quantified using GITT and CV analysis, supported by BVSE mapping that indicates a moderate migration barrier ($\sim$0.76 eV), suggesting efficient Na-ion mobility. To further probe the kinetics, the DRT analysis is employed to deconvolute electrochemical impedance spectra across different time scales. Its excellent agreement with GITT and Warburg-based trends underlines the robustness of the approach. Our detailed analysis provides a deeper understanding of how the material’s structure influences its electrochemical performance, paving the way for its practical use as a promising cathode in future SIB technologies.

\section{\noindent~Experimental}

The Na$_{3.3}$Mn$_{1.2}$Ti$_{0.75}$Mo$_{0.05}$(PO$_4$)$_3$/C cathode material synthesized through sol-gel method using manganese acetate tetrahydrate (Sigma-Aldrich, $\ge$99\%), titanium(IV) bis(ammonium lactato)dihydroxide (Sigma-Aldrich, 50 wt. \% solution in water), ammonium heptamolybdate tetrahydrate (Merck, $\ge$99\%), sodium acetate (Sigma-Aldrich, $\ge$99\%), ammonium dihydrogen phosphate (Thermo-Scientific, $\ge$99.9\%), and citric acid (Sigma-Aldrich, $\ge$99.5\%) as precursors. Here, the citric acid serves dual role: a chelating agent and a carbon source. A stoichiometric amount of precursors was dissolved in de-ionized water, and the aqueous solution was magnetically stirred and heated at 80$\degree$C to obtain a gel, which was further dried at 120$\degree$C overnight in a vacuum oven. The obtained precursors were heated at 650$\degree$C for 10 hrs in a tube furnace under an Argon environment to obtain the final carbon-coated cathode material.

The X-ray diffraction (XRD) patteren (Panalytical Xpert$^3$) was recorded using Cu K$\alpha$ radiation (1.5406 \AA) in the 2$\theta$ range of 10-80\degree. The Raman spectrum was measured with a Renishaw inVia confocal Raman microscope using a 2400 lines/mm grating at a wavelength of 532 nm. The X-ray photoemission spectroscopy (XPS) measurements are done using the AXIS Supra instrument (Kratos Analytical Ltd) utilizing a monochromatic X-ray source of Al-k$\alpha$-1486.6 eV. The field emission scanning electron microscopy (FE-SEM) analysis was performed using a TESCAN Magna LMU equipped with an EDAX AMETEK Octane Elite Super EDS system for elemental mapping. The high-resolution transmission electron microscope images (HR-TEM) and selective area electron diffraction (SAED) pattern were obtained using JEM-ARM200F NEOARM operating at 200 keV.

The electrodes were fabricated by first preparing the slurry of the active material, which was then coated onto battery-grade aluminum foil using the doctor's blade technique. The slurry is composed of active material, conductive carbon, and polyvinylidene fluoride in a ratio of 7:2:1, which is mixed in N-Methyl-2-pyrrolidone. After drying the slurry, 12 mm electrodes were punched and vacuum-dried at 80\degree C before being transferred to the argon-filled glove box (UniLab Pro SP from MBraun). The coin-cells were fabricated using CR2032 casings utilizing prepared electrodes as cathode, sodium metal anode, glass fiber separator, and 1 M NaClO$_4$ in PC with 5\% FEC (v/v) as electrolyte.

The cyclic voltammetry (CV), galvanostatic intermittent titration technique (GITT), and electrochemical impedance spectroscopy (EIS) were performed using a Biologic VMP-3 potentiostat. The galvanostatic charge-discharge (GCD) measurements are taken at the Neware battery analyzer, BTS400. The voltage window for CV, GCD, and GITT measurements is 1.5-4.3 V. For the EIS, the frequency range is from 100 kHz to 10 mHz, and an AC amplitude of 10 mV was used. The EIS spectra were analyzed using the Distribution of Relaxation Time (DRT) method with the help of the open-source MATLAB DRT tools \cite{Wan_EA_15, Plank_JPS_24}. To achieve a reliable fit for the discrete impedance data, Tikhonov regularization was employed along with a nonlinear least-squares fitting approach. The regularization parameter was fixed at 0.0001, using a second-order derivative. Moreover, the full width at half maximum (FWHM) of the radial basis function was set to 0.5 to ensure a reliable fit \cite{Pati_JPS_24}.

\section{\noindent~Results and discussions}

\begin{figure*} 
\includegraphics[width=7.1in]{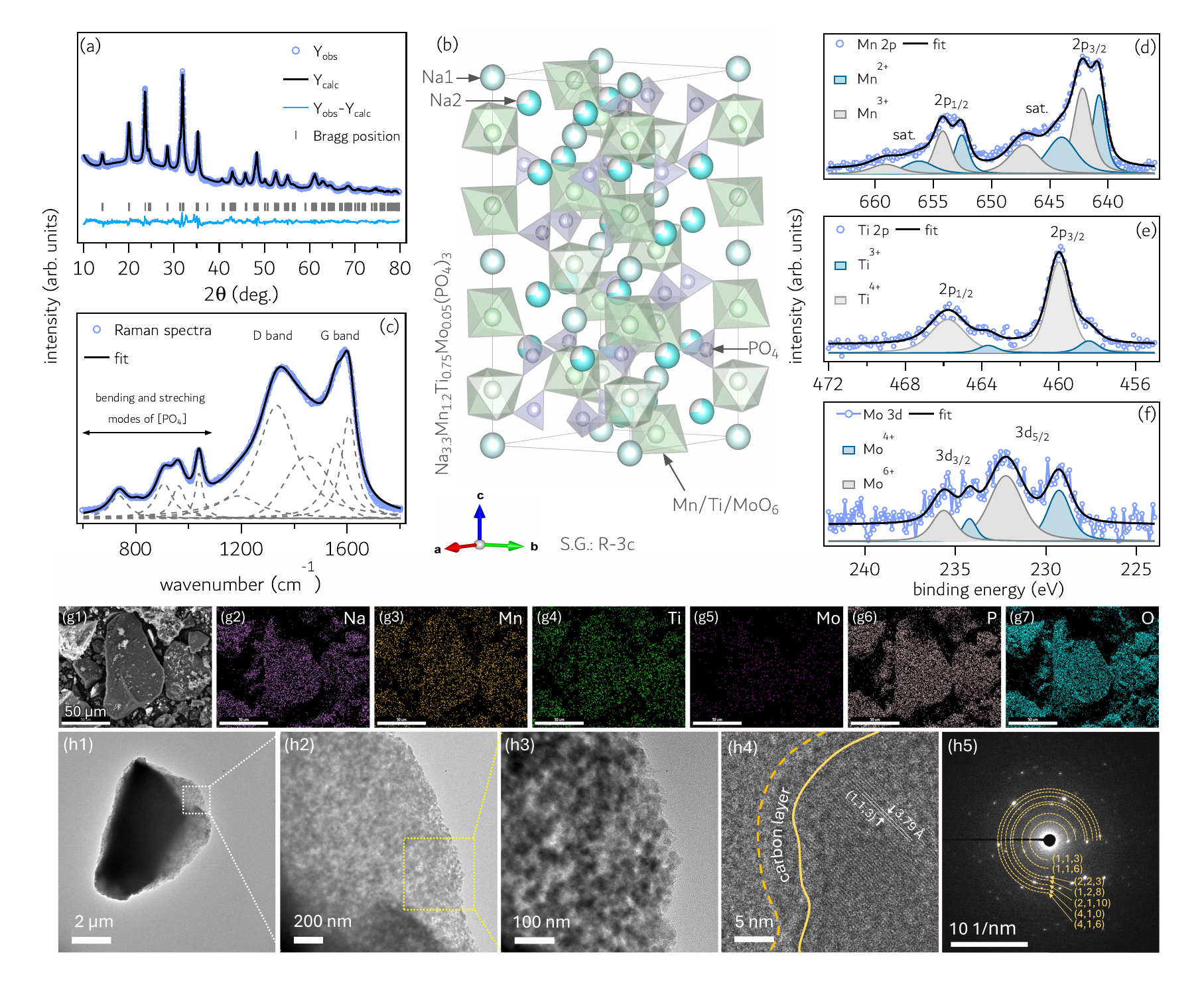}
\caption {The structural, morphological, and compositional analysis of Na$_{3.3}$Mn$_{1.2}$Ti$_{0.75}$Mo$_{0.05}$(PO$_4$)$_3$/C: (a) the Rietveld-refined XRD pattern, (b) the crystal structure, (c) the Raman spectrum, (d–f) room temperature Mn 2$p$, Ti 2$p$, and Mo 3$d$ XPS core-level spectra, (g1) the FESEM image, (g2--g7) the EDS elemental mapping, (h1–h3) the HR-TEM images at different magnifications, (h4) lattice fringes with a measured $d$-spacing and (h5) the SAED pattern.}
\label{Strc}
\end{figure*}

The synthesized Na$_{3.3}$Mn$_{1.2}$Ti$_{0.75}$Mo$_{0.05}$(PO$_4$)$_3$/C cathode material crystallizes in the rhombohedral NASICON-type structure (R$\bar{3}$c space group), as confirmed by the Rietveld refinement of the XRD pattern using FullProf software [see Fig.~\ref{Strc}(a)]. The crystal structure comprises corner-sharing TM octahedra and PO$_4$ tetrahedra forming lantern-like units, which generate an open 3D framework conducive to sodium-ion transport through distinct Na1 and Na2 sites \cite{Manish_PRB_24}, as illustrated in Fig.~\ref{Strc}(b). The refined lattice parameters are determined to be $a$ = $b$ =~8.81~\AA~and $c$ =~21.84~\AA, which are in good agreement with values previously reported in the literature \cite{Gao_CM_16}. These results suggest that the incorporation of a small amount of Mo contributes to stabilizing the increased Mn content (1.2) in the NASICON framework \cite{Liu_JMCA_21, Liu_NE_23}. The detailed atomic coordinates and refinement parameters are listed in Table S1 of \cite{SI}. Moreover, the Raman spectrum shown in Fig.~\ref{Strc}(c) reveals the D and G bands of carbon layers and the vibrational modes of [PO$_4$] unit. The Raman peaks in the range of 700--1100 cm$^{-1}$ belong to the stretching and bending modes of the PO$_4$ unit \cite{Manish_PRB_24, Sapra_ACSAMI_24, Dwivedi_AAEM_21}. The distinct peaks at 1186, 1334, 1451, 1561, and 1606 cm$^{-1}$ correspond to the D4, D1, D3, G, and D2 bands, respectively, which characterize the amorphous carbon coating. The D1 band (1334 cm$^{-1}$) arises from disorder-induced vibrations, indicating structural disorder within the carbon coating, while the D3 band (1451 cm$^{-1}$) is associated with amorphous carbon. The D4 band (1186 cm$^{-1}$) corresponds to the stretching vibrations of sp$^2$-sp$^3$ bonds. The G band (1561 cm$^{-1}$) attributed to the E$_{2g}$ vibrational mode of graphite, suggests the presence of small crystalline graphitic domains within the disordered matrix. Additionally, the D2 band (1606 cm$^{-1}$) is analogous to the G band but represents the vibrational modes of surface graphene layers. This indicates the presence of both disordered and graphitic domains within the amorphous carbon coating, which enhances surface electronic conductivity and supports efficient charge transport in the active material \cite{Sadezky_C_05, Sharma_IJPAP_24}. 

\begin{figure*} 
\includegraphics[width=7in]{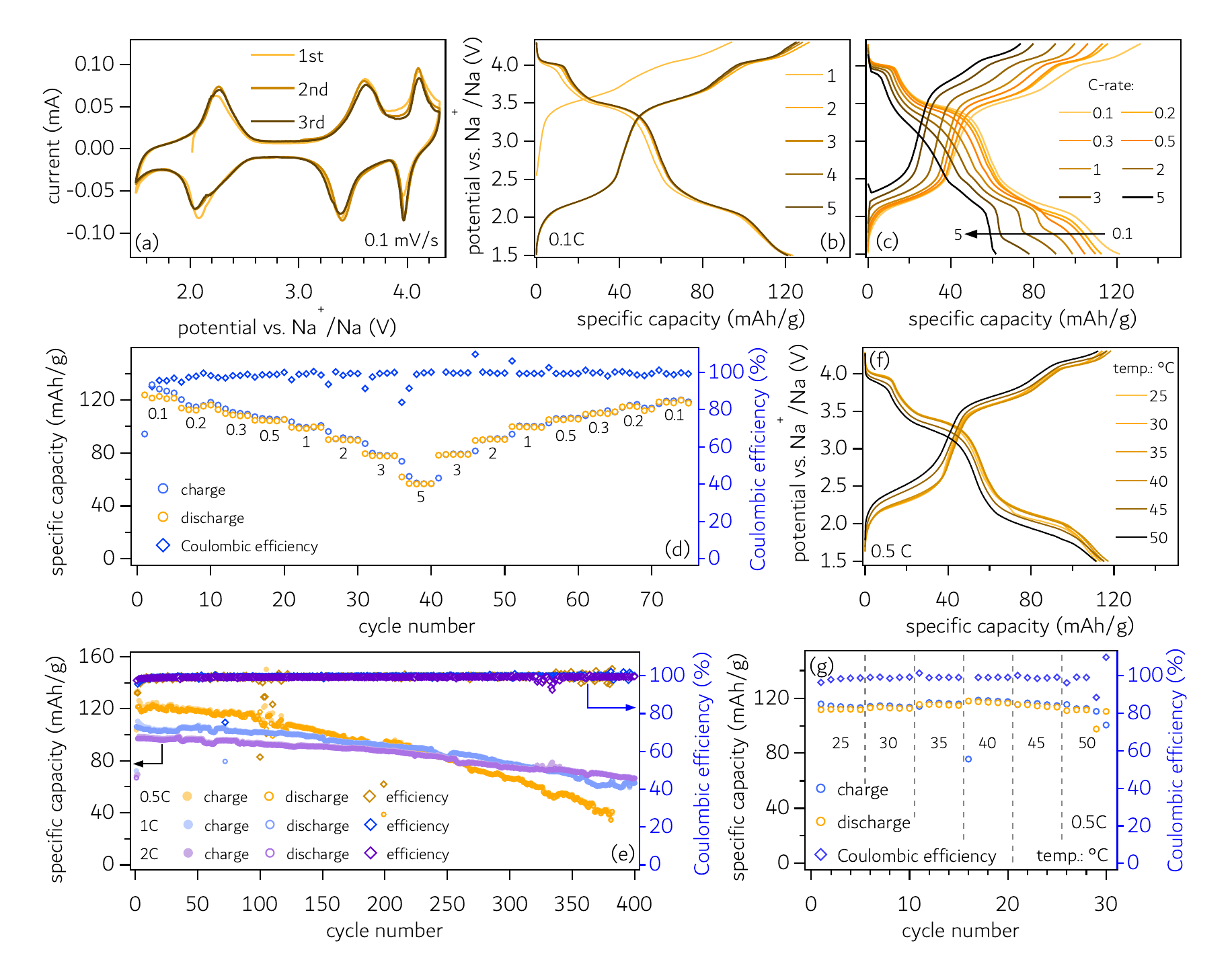}
\caption {The electrochemical performance of Na$_{3.3}$Mn$_{1.2}$Ti$_{0.75}$Mo$_{0.05}$(PO$_4$)$_3$/C cathode material in the potential window of 1.5–4.3 V: (a) initial CV curves at 0.1 mV/s, (b) first five GCD profiles at a current rate of 0.1 C, (c) second GCD cycle at various current rates, (d) the rate capability performance at various C-rates, (e) the long-term cycling stability tests at 0.5, 1, 2 C. (f) The GCD profiles in the second cycle, measured at 0.5 C, shown at selected temperatures and (g) the cyclic performance.}
\label{CV_GCD}
\end{figure*}

The XPS measurements are performed to analyze the surface composition and chemical states of the pristine sample where the survey spectrum shown in Fig.~S1(a), confirms the presence of all constituent elements. The corresponding core-level spectra of Na 1$s$, P 2$p$, O 1$s$, and C 1$s$ are presented in Fig.~S1(b-e). All the core-level spectra are calibrated using the C 1s peak at 284.6 eV as a reference. For the XPS analysis, we first subtracted the inelastic background following the Tougaard method, and then fitted the core-levels (in IGOR Pro software) using the Voigt function, which includes both the Gaussian and Lorentzian profiles \cite{Manish_PRB_24}. The core level of Mn 2$p$ shown in Fig.~\ref{Strc}(d) reveals two strong peaks belonging to spin-orbit split 2$p_{3/2}$ and 2$p_{1/2}$ states along with the broad satellites. Through  the deconvolution, we find that the characteristic 2$p_{3/2}$ peak splits into two components located at 640.8 and 642.2 eV, and so is the 2$p_{1/2}$ peak at 652.6 and 654.2 eV \cite{Zhu_AEM_19}. These split peaks can be assigned to the 2+ and 3+ oxidation states of Mn present at the surface of the pristine sample in almost equal ratios (calculated using the integral area under the peaks) \cite{Zhu_AEM_19}. In contrast, the Ti 2$p$ core level spectrum [see Fig.~\ref{Strc}(e)] shows the 2$p_{3/2}$ characteristic peak located at 460.0 eV and 2$p_{1/2}$ at 465.8 eV, which is the signature of the Ti$^{4+}$ state along with a little contribution of Ti$^{3+}$ located at 458.4 eV (2$p_{3/2}$) and 463.7 eV (2$p_{1/2}$) \cite{Li_MTE_21}. The Mo 3$d$  core-level presented in Fig.~\ref{Strc}(f) exhibits characteristic peaks for the Mo$^{6+}$ oxidation state, with the 3$d_{5/2}$ and 3$d_{3/2}$ components appearing at 232.2 eV and 235.6 eV, respectively. Additionally, peaks corresponding to the Mo$^{4+}$ oxidation state are observed at 229.3 eV (3$d_{5/2}$) and 234.2 eV (3$d_{3/2}$). This indicates the presence of a mixed-valence state of the dopant Mo, where the intensity ratio suggests that one-third of Mo is reduced to Mo$^{4+}$ state \cite{Tan_NR_22}. 

The FE-SEM image presented in Fig.~\ref{Strc}(g1) reveals agglomerated particles with irregular morphology, while Fig.~S2(a) of \cite{SI} indicates a random particle size distribution within the submicron range. The observed rough surface texture may help in better electrolyte penetration during electrochemical processes. The elemental distribution across the same region in Fig.~\ref{Strc}(g1), as shown in the EDS maps in Fig.~\ref{Strc}(g2-g6), confirms a homogeneous dispersion of Na, Mn, Ti, Mo, P, and O elements. The corresponding EDS mass sum spectrum is shown in Fig.~S2(b) of \cite{SI}. The HR-TEM images provide valuable insight into the structural characteristics of the synthesized sample. Initially, the micrograph in Fig.~\ref{Strc}(h1) reveals micron-sized particles as predicted from the FESEM images. However, upon examining the expanded region within the white box [see Fig.~\ref{Strc}(h2)], the material appears predominantly amorphous, exhibiting variations in contrast density. A more detailed investigation of the confined area highlighted by the yellow box in Fig.~\ref{Strc}(h2) reveals the presence of nanocrystallites, which appear as darker spots embedded within the amorphous carbon matrix, as shown in Fig.~\ref{Strc}(h3). At even higher magnification, distinct lattice planes become visible, confirming the existence of nanocrystalline domains enclosed within the carbon matrix [see Fig.~\ref{Strc}(h4)]. The average crystallite size estimated from the HR-TEM image is approximately $\sim$18.1 nm (see Fig.~S3 of \cite{SI}), which aligns well with the value obtained from XRD analysis using the Scherrer equation, yielding an average size of 20.3 nm. Additionally, the SAED pattern presented in Fig.~\ref{Strc}(h5) exhibits bright spots corresponding to the (1,1,3), (1,1,6), (2,2,3), (1,2,8), (2,1,10), (4,1,0), and (4,1,6) crystal planes. These planes have interplanar spacings of 0.398, 0.278, 0.212, 0.198, 0.175, 0.159, and 0.149 nm, respectively, aligning well with the results obtained from the XRD analysis.

After building on the obtained structural insights, we aim to evaluate the electrochemical performance of the material to understand its sodium storage behavior and diffusion kinetics. Therefore, first we  discuss the intial three cyclic voltammetry (CV) curves taken at a scan rate of 0.1 mV/s in a half-cell configuration, depicted in Fig.~\ref{CV_GCD}(a), which show three peaks each in anodic/cathodic cycle at voltages of 2.26/2.06, 3.60/3.40, and 4.10/3.98 V, belonging to the redox couples of Ti$^{3+}$/Ti$^{4+}$, Mn$^{2+}$/Mn$^{3+}$, and Mn$^{3+}$/Mn$^{4+}$, respectively \cite{Liu_JMCA_21}. We find no peaks in CV at 4.0 V during charge or 2.5 V during discharge, indicating the absence of IASD related to Mn hopping \cite{Zhang_ACSEL_21}. Moreover, the overlapping of successive CV curves and the consistent peak current values indicate good reversibility of the electrochemical processes. Further, the first five galvanostatic charge-discharge (GCD) cycles measured at 0.1 C (1 C~=~114 mAh/g) are shown in Fig.~\ref{CV_GCD}(b) where the initial charge reveals no contribution from the Ti$^{3+}$/Ti$^{4+}$ redox, and two-step features at average voltages of 3.6 and 4.1 V belong to the Mn$^{2+}$/Mn$^{3+}$, and Mn$^{3+}$/Mn$^{4+}$, respectively. During the discharge, both the Mn redox are reversibly obtained along with the appearance of Ti$^{3+}$/Ti$^{4+}$ redox at a voltage level of 2.1 V \cite{Zhu_AEM_19}, and a discharge capacity of 124 mAh/g is achieved, which found to be better than other reported Mn-based ternary NASICON-type cathodes \cite{Jiang_ESM_24, Zhu_ASS_24, Ren_CEJ_24}, as listed in Table S2 of \cite{SI}, considering the fact of higher Mn content in the present case. After the initial cycle, the subsequent charge/discharge profiles reciprocate and retain 98\% of the initial discharge capacity after five cycles, which signifies stable electrochemical activity. However, if we look at the GCD profile, the plateaus seem to be sloppy, which is a signature of the dominant electrochemical contributions from the surface states compared to the bulk states, as expected from the carbon-coated nano-crystallite morphology of the sample confirmed using HR-TEM images \cite{Okubo_JACS_07}.

\begin{figure} 
\includegraphics[width=3.4in]{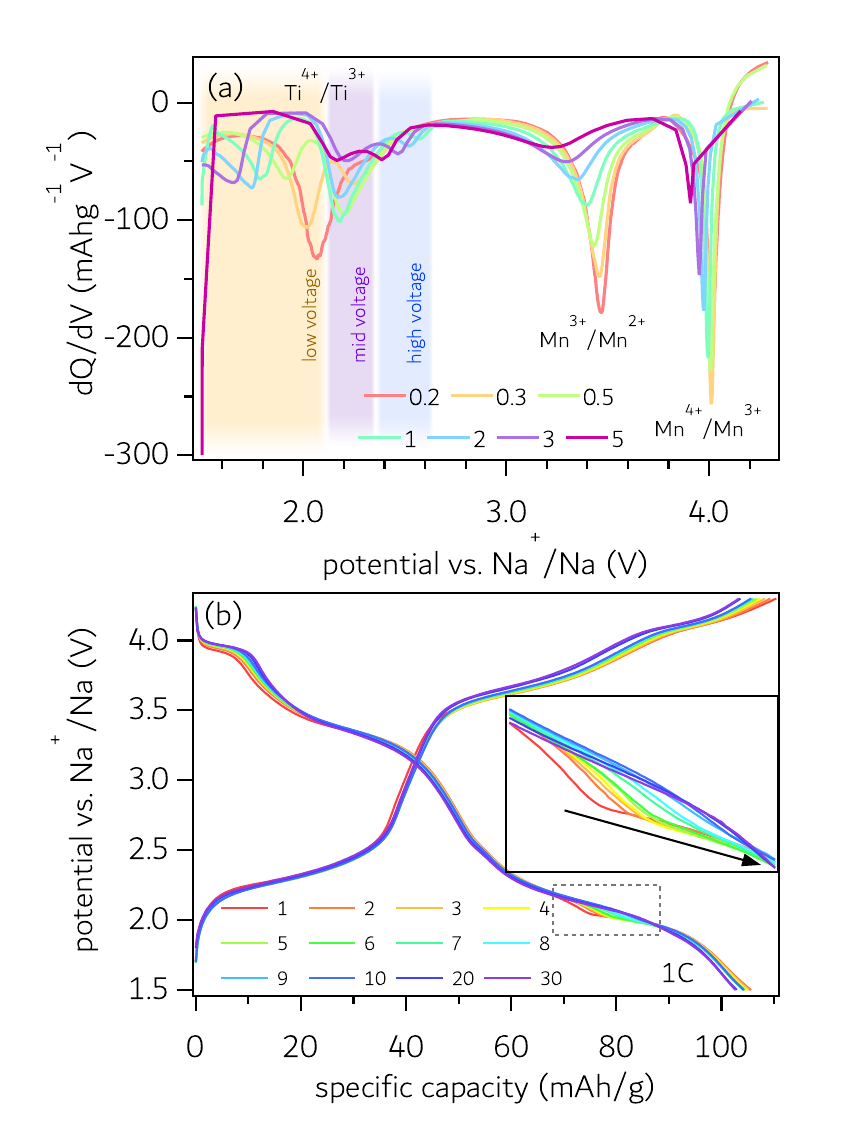}
\caption {(a) The dQ/dV curves of the discharge profiles [shown in Fig.~\ref{CV_GCD}(c)], (b) the GCD profiles measured at 1 C up to 30 cycles, and inset shows the zoomed view of the dashed region.}
\label{dQ/dV}
\end{figure}

Moreover, the GCD profiles at various current rates between 0.1 and 5 C are presented in Fig.~\ref{CV_GCD}(c), indicating stable electrochemical behavior. The corresponding specific discharge capacities in the second cycle are found to be 121.4, 112.6, 109.8, 104.7, 98.7, 90.7, 77.8, and 61.8 mAh/g, respectively, demonstrating good rate capability [see Fig.~\ref{CV_GCD}(d)], along with efficient recovery of specific capacity upon reverting to lower C-rates. The long cycling performance evaluated at 0.5, 1, and 2 C showed capacity retentions of 55\%, 73\%, and 77\%, respectively, after 300 cycles [see Fig.~\ref{CV_GCD}(e)]. Notably, at 2 C, approximately 70\% of the capacity is retained even after 400 cycles. We find that the average capacity fade per cycle is high at lower C-rates, whereas capacity retention improves at higher C-rates. This behavior can be ascribed to the shorter exposure time for the electrode remaining at high potentials during faster cycling, thereby suppressing undesirable interfacial reactions \cite{Pati_JMCA_22}. More interestingly, the GCD profiles measured at elevated temperatures with a current rate of 0.5 C [see Fig.~\ref{CV_GCD}(f)] show no notable change in the capacity values up to 50\degree C. This suggests that the charge storage process is not limited by electrode kinetics but rather influenced by the sample morphology \cite{Okubo_JACS_07, Wu_JMC_11}. The GCD profiles are found to be nearly identical up to 40\degree C, but a significant increase in polarization between the charge and discharge curves is visible at higher temperatures. This may be due to enhanced interface-electrolyte catalytic reactions at higher temperatures, which could lead to the formation of a thicker solid-electrolyte interface (SEI) and cause a hindrance to the ionic motion \cite{Dong_JEC_22}. Further, the cycling tests at different temperatures up to 30 cycles presented in Fig.~\ref{CV_GCD}(g) confirm the thermal and cyclic stability of the cathode material. 

\begin{figure*} 
\includegraphics[width=7.3in]{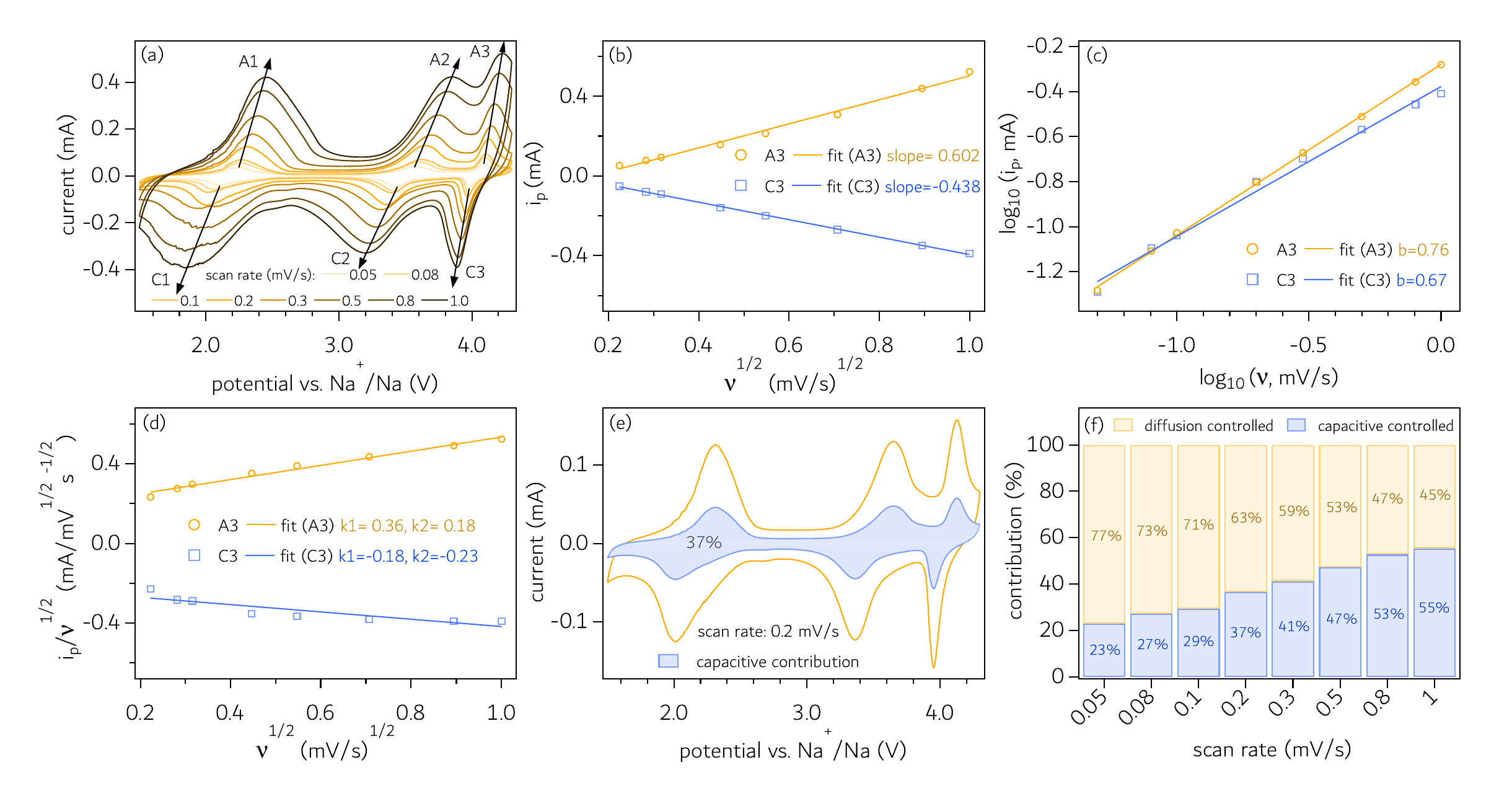}
\caption {The electrochemical characterization of Na$_{3.3}$Mn$_{1.2}$Ti$_{0.75}$Mo$_{0.05}$(PO$_4$)$_3$/C at various scan rates (0.05–1.00 mV/s) in the potential window of 1.5--4.3 V: (a) The CV curves at different scan rates, (b) the linear relationship between peak current (i$_p$) and the square root of the scan rate ($\nu^{1/2}$), (c) the logarithmic correlation between log(i$_p$) and log($\nu$), (d) a linear fitting of i$_p$/$\nu^{1/2}$ versus $\nu^{1/2}$ for both anodic and cathodic peaks, (e) the capacitive contribution at 0.2 mV/s, highlighted by the shaded area and (f) the contributions of the capacitive and diffusive contributions to the total current at varying scan rates.}
\label{CV}
\end{figure*}

Note that during discharge, a step-like voltage drop is observed in the GCD profile after a voltage of 2.15 V, and the extent of the voltage drop increases with increasing current rates, see Fig.~\ref{CV_GCD}(c). To investigate this observation further, the dQ/dV curves obtained from the discharge profiles are shown in Fig.~\ref{dQ/dV}(a). The redox region of Ti$^{3+}$/Ti$^{4+}$ shows no peak split at a lower current rate of 0.2 C. However, at 0.3 C and above, a significant splitting is visible, which increases with the current rate. Moreover, no such splitting is observed in peaks belonging to the Mn redox. Such voltage steps could be produced by an increase in polarization of the passivation layer/SEI on the sodium counter electrode instead of a change in the potential of the working electrode \cite{Rudola_EC_14}. However, this voltage step gradually diminishes with continued GCD cycling and eventually disappears after several cycles at 1 C, as shown in Fig.~\ref{dQ/dV}(b) up to 30 cycles. This behavior may be attributed to the stabilization of the passivation layer. A small peak in the shaded high-voltage region shows a monotonic voltage shift with increasing C-rate, similar to the Mn redox peaks. This feature around 2.5 V likely indicates Mn hopping, consistent with previous reports \cite{Zhang_ACSEL_21, Liu_NE_23}. It shows small intensity compared to other redox peaks suggests that Mo-doping effectively suppresses the Mn hopping \cite{Zhu_AEM_19, Liu_NE_23}.

Now we move to understand the charge storage mechanism in electrode material, whcih can arise from both faradic and non-faradic processes. The faradic processes involve diffusion-controlled alkali metal ion intercalation and associated redox reactions, while non-faradic processes are primarily surface-controlled include pseudocapacitance and the double-layer effect  \cite{Fleischmann_CR_20, Gogotsi_ACSN_18}. In order to get a comprehensive idea and to quantify these contributions, a detailed analysis of the CV data, as shown in Fig.~\ref{CV}(a), is performed at various scan rates from 0.05--1 mV/s in the potential window of 1.5-4.3 V. We observe a shift in the oxidation peaks towards higher potentials and the reduction peaks towards lower potentials when the scan rate increases. This shift is linked to an increase in polarization at higher current/scan rates. Additionally, the magnitudes of both oxidation and reduction peaks become more pronounced with increasing scan rates, which can be attributed to the presence of electric double-layer capacitors, with the capacitive current enhanced with the scan rate \cite{Gogotsi_ACSN_18, Malik_JES_24}. Notably, the voltage drop obtained in the GCD profile is not significant in the CV curves; however, fluctuations are visible in that particular voltage range ($\le$$\sim$2.15 V), whereas the rest of the voltammograms remain smooth. The discrepancy in polarization behavior observed between GCD and CV data could be attributed to the different response of the passivation layer under constant current (GCD) and dynamic potential (CV) conditions; however, further investigation is required to determine the underlying cause precisely. 

To probe the kinetic behavior, we calculate the diffusion coefficient values for both cathodic and anodic peaks using the Randles-\v{S}ev\v{c}ik equation \cite{Sharma_IJPAP_24, Manish_CEJ_23}.
\begin{equation}
i_p = (2.69 \times 10^5) A D^{\frac{1}{2}} C\eta^{\frac{3}{2}} \upsilon^{\frac{1}{2}}
\label{RS}
\end{equation}
where $i_p$ (mA) is the peak current, $A$ (cm$^2$) is the electrode surface area, $D$ (cm$^2$/s) is the diffusion coefficient, $C$ (mol/cm$^3$) is the concentration of Na ions in bulk, $\eta$ is the number of transferred electrons in the redox reaction, and $\upsilon$ (mV/s) is the applied scan rate. The peak current of A3 and C3 obtained from the CV curves is plotted as a function of the square root of the scan rate [see Fig.~\ref{CV}(b)], and the slope obtained from the linear fit along with equation \ref{RS} provided the diffusion coefficient in the range of 2--5$\times$10$^{-10}$ cm$^2$/s. However, we note that the Randles-\v{S}ev\v{c}ik equation accounts for the Faradic processes and overlooks any contributions from pseudocapacitance. On the other hand, the observed broad CV peaks and sloppy GCD profiles suggest a combination of Faradaic and pseudocapacitive behavior \cite{Okubo_JACS_07, Malik_JES_24}. To qualitatively assess the contributions at various scan rates, we present the log($i$) versus log($\upsilon$) plot in Fig.~\ref{CV}(c) and conduct a linear fit using the power law \cite{Wang_JPCC_07}. 
\begin{equation}
i = a \upsilon^b
\label{power_law}
\end{equation}
\begin{figure*} 
\includegraphics[width=7.2in]{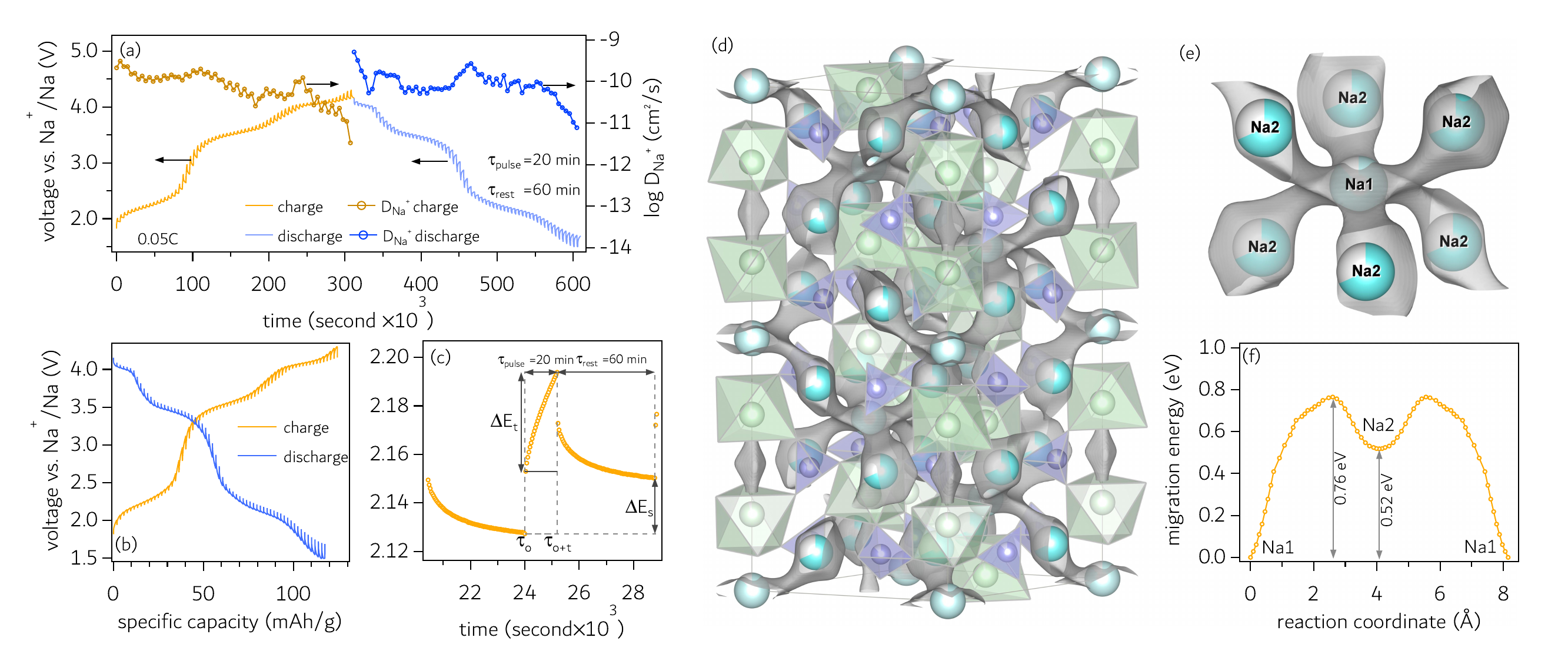}
\caption {The GITT measurement at a current rate of 0.05 C in a voltage window of 1.5–4.3 V plotted vs. (a) time and (b) specific capacity, after 5 GCD cycles at 0.05 C, (c) the schematic labeling of different parameters of a single titration curve before, during and after application of a current pulse for 20 min, (d) the calculated iso-surfaces for sodium-ion 3D migration channels (iso-surface level~=~0.7 eV), (e) the connectivity of the Na2-Na1-Na2 migration channel, and (f) the calculated activation energy of the sodium-ion migration.}
\label{Diffusion}
\end{figure*}
here $i$ is the current, $\upsilon$ is the scan rate, and $a$ and $b$ are the fitting parameters. The value of $b$ determines the nature of the sodium ion charge/discharge reaction mechanism. If $b$ value is $\le$0.5, the storage process is semi-infinite diffusion-controlled and reduces to the RS equation, while for $b=$~1, the process is a surface-limited capacitive process \cite{Gogotsi_ACSN_18, Malik_JES_24}. The collective contribution from both processes leads to the intermediate value of $b$, marking the finite-length diffusion. The obtained $b$ values are 0.76 and 0.67 for anodic (A3) and cathodic (C3) peaks, respectively, and are indicative of mixed control of the charge storage process. Therefore, in order to quantify these, the peak current can be written as a linear combination of surface and diffusion-controlled reactions, as proposed and adopted in refs.~ \cite{Liu_JES_98, Wang_JPCC_07}: 
\begin{equation}
\frac{i}{\upsilon^{\frac{1}{2}}} = k_1 \upsilon^{\frac{1}{2}} + k_2
\end{equation}
where $k_1$ and $k_2$ are the coefficients for the respective capacitive and diffusion currents, which are calculated by linear fitting of the $i$/$\upsilon^{1⁄2}$ versus $\upsilon^{1⁄2}$, as shown in Fig.~\ref{CV}(d), the values for $k_1$/$k_2$ are found to be 0.36/0.18, and 0.18/0.23 for the A3 and C3 peaks, respectively. These values indicate that the diffusion part is dominant in cathodic reactions and has a high capacitive contribution in the case of an anodic reaction. The $k_1$ and $k_2$ are further utilized to quantify the diffusive and capacitive contribution as a function of scan rate, and the shaded region in Fig.~\ref{CV}(e) shows that around 37\% of the charge storage contribution is capacitive at 0.2 mV/s. Likewise, at a lower scan rate of 0.05 mV/s, the diffusion part is dominant with 77\% contribution [see Fig.~\ref{CV}(f)] and gradually decreases with increment in scan rates, indicating a major influence of pseudocapacitance and surface-controlled reactions on the charge-storage process at higher scan rates \cite{Li_ESM_20}. Note that the $k_1$ and $k_2$ based analyses mentioned above fail to consider the shifts in redox peak potentials caused by the polarization effects, which become increasingly prominent at higher sweep rates. A similar limitation applies to the above discussed Randles-\v{S}ev\v{c}ik equation, which also considers the redox peak currents, making the diffusion coefficient inclusive of charge transfer kinetics and pseudocapacitive effects. 

\begin{figure*}
\includegraphics[width=7.3in]{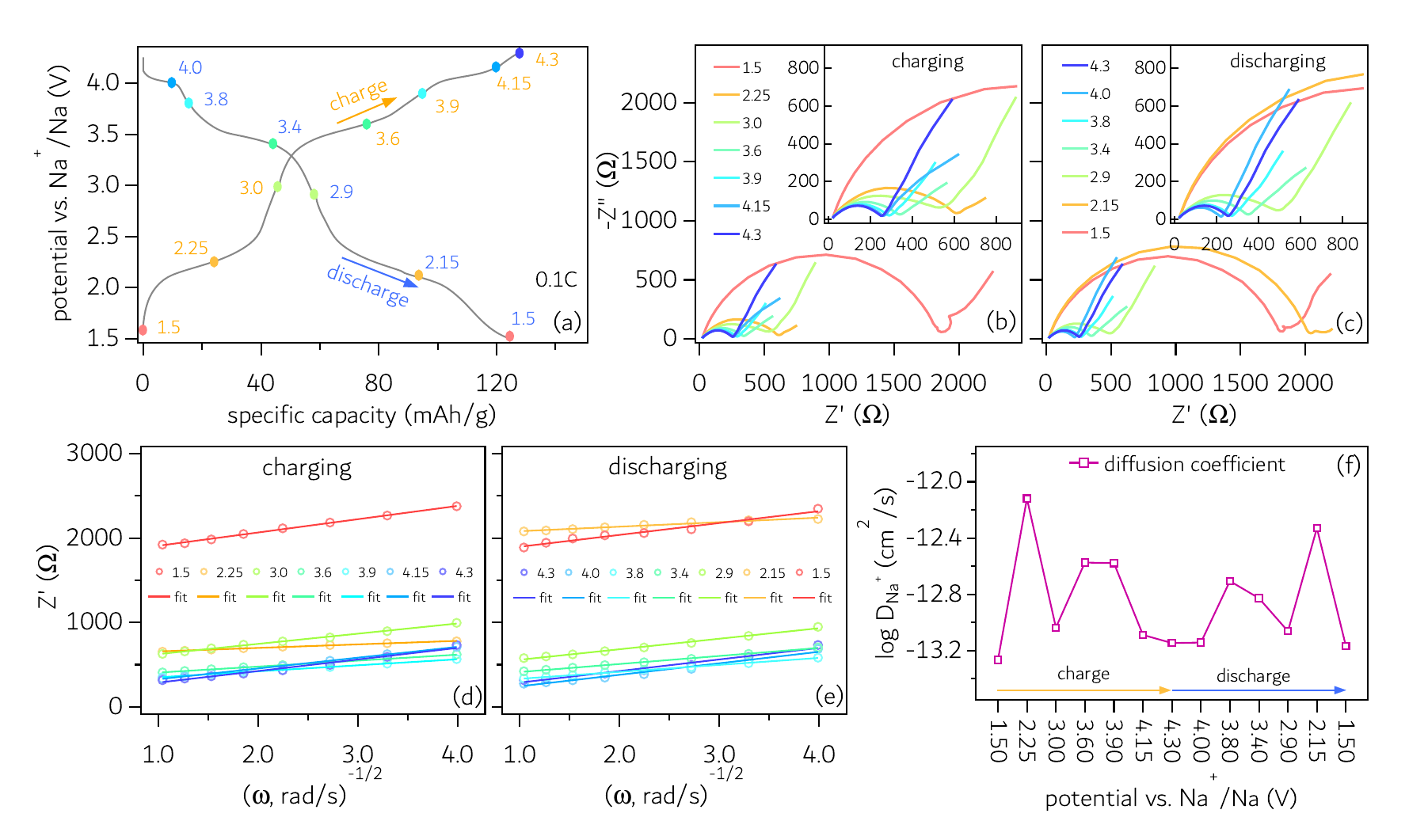}
\caption {(a)The GCD profile with points marked at the particular voltage for the EIS measurements. The EIS data in the Nyquist plot during (b) charging, and (c) discharging (inset shows the zoomed view). A linear fit of the real impedance versus the inverse of the square root of angular frequency in the Warburg region during (d) charging, and (e) discharging, (f) the diffusion coefficient at various voltages calculated using the Warburg coefficient obtained from Fig.~\ref{EIS}(d,e)}
\label{EIS}
\end{figure*}

Note that to calculate the diffusion coefficient, the GITT method is generally considered more accurate and reliable because it operates under near-equilibrium conditions, minimizing kinetic and capacitive effects. Therefore, the GITT measurements are performed within the voltage range of 1.5--4.3 V at a rate of 0.05 C. Before the measurement, the freshly fabricated cell was given five formation charge/discharge cycles to stabilize the electrode. The time-resolved GITT profile is shown in Fig.~\ref{Diffusion}(a). The cell was initially charged to the upper cutoff voltage of 4.3 V using a 20 mins current pulse ($\tau_{\text{pulse}}$), followed by a 60 mins relaxation period to allow uniform sodium-ion distribution and to reach a quasi-equilibrium open-circuit voltage (OCV), where the dE/dt (V/s)$\rightarrow$0. This charge-relaxation sequence was continuously repeated during discharge until the lower cutoff voltage of 1.5 V was reached. The corresponding GITT voltage profile as a function of capacity is presented in Fig.~\ref{Diffusion}(b).
To estimate the diffusion coefficient under thermodynamic equilibrium, the Fick’s second law is simplified using some assumptions, including a semi-infinite diffusion model, as below \cite{Nickol_JES_20}: 
\begin{equation}
	D_{\text{Na}^+} =\frac{4}{\pi \tau }\left(\frac{m_{B}V_{M}}{M_{B\ }A}\right)^2\left(\frac{{\Delta E}_S}{\tau \left(\frac{dE_{\tau }}{d\sqrt{\tau }}\right)}\right)^2\ \  ; \tau \ll L^2/D_{{Na}^+}  
\label{FICK1}
\end{equation}
here, $m_B$ and $M_B$ denote the active mass (in g) and molecular weight (in g/mol) of the cathode material, respectively. The $V_M$ is molar volume (cm$^3$/mol), $L$ is the electrode thickness ($\mu$m), and $A$ represents the active surface area of the electrode (cm$^2$). The $\tau$ is the duration of the applied current pulse (s), ${\Delta}E_S$ is the potential change before and after the pulse, and the ${\Delta}E_\tau$ is the voltage difference between the equilibrium potential and the peak potential at the end of the current pulse as shown in Fig.~\ref{Diffusion}(c). Assuming a linear relationship between the transient voltage and ${\tau}^{1/2}$, the above equation can be simplified as follows:
\begin{equation}
	D_{\text{Na}^+} =\frac{4}{\pi \tau }\left(\frac{m_{B}V_M}{M_{B\ }A}\right)^2\left(\frac{{\Delta E}_S}{\Delta E_{\tau }}\right)^2 
\label{FICK2}
\end{equation}
The apparent Na$^+$ diffusion coefficient values ($D_{\text{Na}^+}$) of the cathode are estimated to be in the range of $10^{-9.5}$ to $10^{-10.5}$~cm$^2$/s across the voltage window of 4.3--1.5~V for both charge and discharge processes, as shown in Fig.~\ref{Diffusion}(a). These values are comparable to those reported for other high-performance cathode materials and are crucial for the good rate capability demonstration \cite{Li_ESM_20, Li_AEM_18}. Notably, in the Ti$^{3+}$/Ti$^{4+}$ redox region, the $D_{\text{Na}^+}$ remains relatively stable between $10^{-9}$ and $10^{-10}$~cm$^2$/s. In contrast, within the Mn$^{2+}$/Mn$^{3+}$ region, the $D_{\text{Na}^+}$ fluctuates just below $10^{-10}$~cm$^2$/s and further decreases in the Mn$^{3+}$/Mn$^{4+}$ region, indicating more sluggish Na$^+$ transport and slower reaction kinetics of Mn$^{2+}$/Mn$^{3+}$/Mn$^{4+}$ redox than Ti$^{3+}$/Ti$^{4+}$, which is consistent with earlier observations in  \cite{Li_ESM_20}. Further, we use SoftBV GUI tool to evaluate the sodium-ion migration behavior within the bulk of the cathode material \cite{Chen_ACB_19}. The SoftBV utilizes the bond valence method to calculate the energy landscape and generate iso-surfaces corresponding to possible sodium-ion migration pathways. The crystallographic information file, obtained from Rietveld refinement of the XRD pattern is used as input for the simulation. The results predict the most favorable sodium-ion migration pathways, with the 3D iso-surfaces of these diffusion channels, as illustrated in Fig.~\ref{Diffusion}(d). These pathways connect the Na1 and Na2 sites via Na1--Na2--Na1 or Na2--Na1--Na2 sequences, indicating continuous and interconnected migration channels. The corresponding channel topology is visualized in Fig.~\ref{Diffusion}(e). The estimated activation energy barrier for sodium-ion migration is found to be 0.76 eV, as shown in Fig.~\ref{Diffusion}(f), which is consistent with values reported in \cite{Li_ESM_20, Chen_S_24, Gokulnath_EF_24}. 

\begin{figure*}
\includegraphics[width=6.8in]{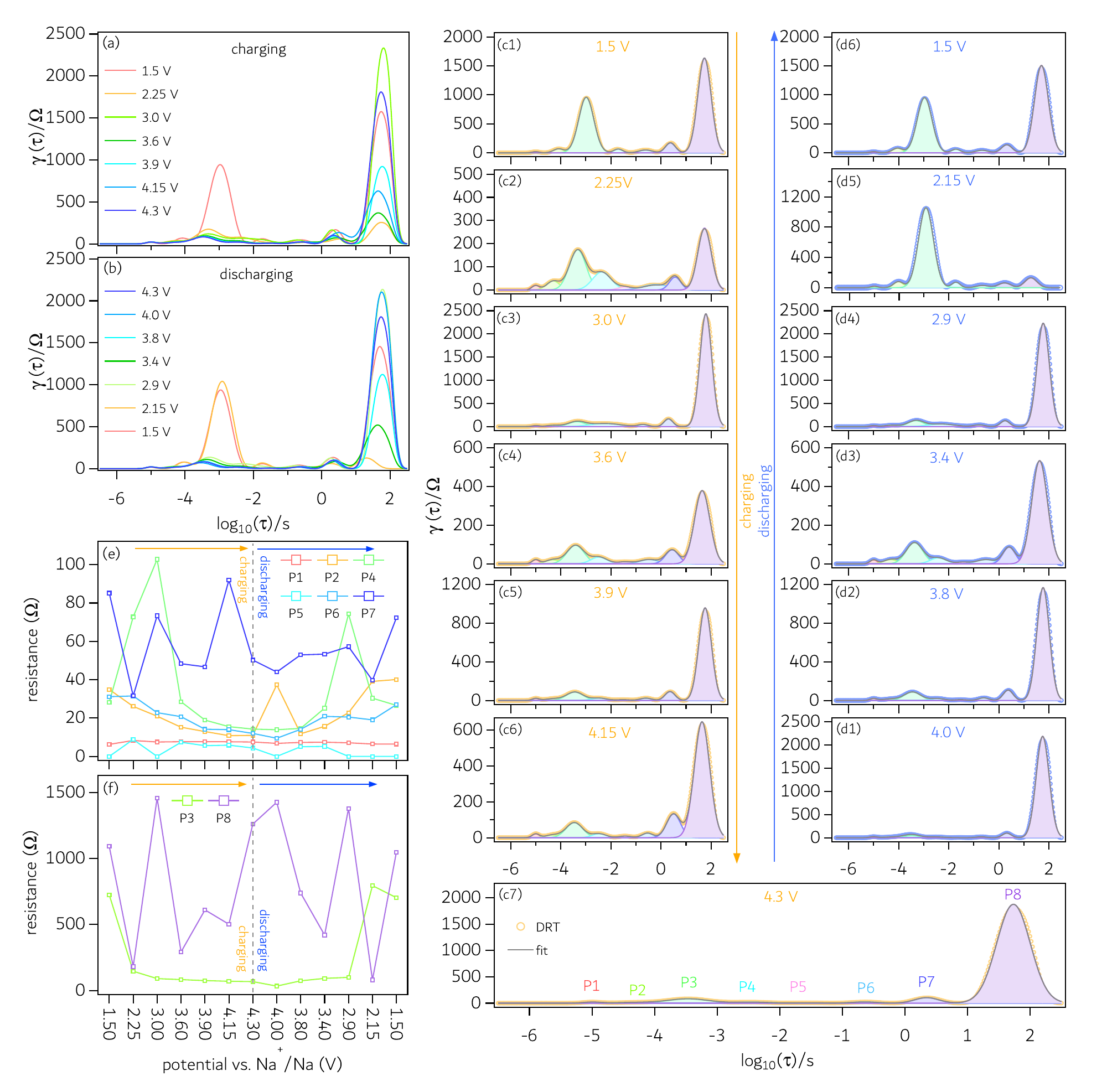}
\caption {The DRT curves, simulated from the EIS spectra at various voltages (a, b) in the voltage range of 1.5-4.3 V, and the fitted DRT spectra during (c1-c7) charging and (d1-d6) discharging, respectively. The resistance values calculated from the peak fitting of the DRT curves are plotted as a function of voltage: (e) P1, P2, P4, P5, P6, P7, and (f) P3, P8.}
\label{DRT}
\end{figure*}

After analyzing the sodium-ion kinetics through GITT and evaluating the activation energy barriers, further insight into the electrode-electrolyte interfacial processes is vital to understand different reaction mechanisms. Therefore, we recorded the EIS data at specific voltages indicated in Fig.~\ref{EIS}(a), corresponding to different stages of de/sodiation and the corresponding Nyquist plots are shown in Figs.~\ref{EIS}(b, c), respectively. Prior to recording the EIS spectrum, we stabilize the electrode-electrolyte interface by performing charge-discharge at 0.1 C up to 3 cycles. At first glance, the Nyquist plots appear similar across different states, featuring a depressed semicircle attributed to the combined effects of SEI, charge transfer resistance, and other electrochemical processes, followed by a straight-line region representing solid-state diffusion (Warburg impedance). Throughout various charge/discharge states, the resistance from all electrochemical processes (except Warburg impedance) remains relatively low (200--600 $\Omega$), except at the lower cut-off voltage of 1.5 V, where it increases around 1900 $\Omega$. Interestingly, a sharp increase in impedance is seen at 2.15 V during discharge, which is the active region for the Ti$^{3+}$/Ti$^{4+}$ redox couple, and could be related to the voltage drop we found in the GCD profiles. Although this might seem surprising, it is consistent with our GITT findings, which indicate enhanced sodium-ion kinetics in the same region. This suggests that the observed impedance rise could be because of a different electrochemical process that does not restrict the sodium-ion diffusion in bulk. To deconvolute the SEI and charge transfer contributions, the Nyquist plots are fitted, as shown in Figs.~S4(a1--a7) and (b1--b6) of \cite{SI} during the charge and discharge, respectively. The impedance spectra are validated using the Kramer-Kronig test performed with LIN-KK software \cite{Schonleber_EA_14}, and the corresponding residuals are presented in Figs.~S4(c1--c13) of \cite{SI}. Here, the equivalent circuit in Fig.~S4(d) is used for the EIS fitting of all the spectra using EC-lab V11.60 software. The R1, R2, and R3 resistive elements belong to the solution/ohmic, SEI, and charge transfer resistances, respectively. Whereas, the Q1, Q2, and Q3 elements stand for the constant phase element belonging to the SEI, charge transfer, and Warburg region, respectively. The resistance values R1, R2, and R3 across different voltages are shown in Fig.~S5 of \cite{SI}. The solution resistance R1 stays fairly constant during operation, with a slight drop only at the lower cut-off voltage of 1.5 V. On the other hand, either of the R2 and R3 values is difficult to link to the sodium-ion diffusion resistance. Consequently, this analysis doesn't offer any definitive information about the electrochemical behavior \cite{Malik_JES_24}. However, the diffusion kinetics of Na-ions inside the bulk of the electrode at different voltages can be estimated by analyzing the Warburg region of the EIS data using the following equations \cite{Manish_CEJ_23}:
	\begin{equation}
	D_{Na^+}=\frac{R^2 T^2}{2A^2\eta^4F^4C^2\sigma^2}
	\label{diffusion}
	\end{equation}
	\begin{equation}
	Z'=R_S +R_{CT}+\sigma\omega^{-0.5}
	\label{warburg}
	\end{equation}
here, R represents the gas constant, T is the temperature, A is the geometrical surface area of the electrode, $\eta$ is the number of sodium ions involved in the reaction, F is the Faraday constant, C denotes the Na concentration in the electrode material (mol-cm$^3$), and $\sigma$ is the Warburg coefficient. The Warburg coefficient is extracted using equation~\ref{warburg}, in which Z$'$ is the real part of the impedance, R$_s$ and R$_{ct}$ are the solution and charge transfer resistance, respectively. 
The real part of impedance (Z$'$) is plotted as a function of the inverse square root of the angular frequency ($\omega^{-0.5}$), and the data points are fitted using equation~\ref{warburg}, as shown in Figs.~\ref{EIS}(d, e). Using the parameters in equation~\ref{diffusion}, the calculated $D_{\text{Na}^+}$ values at OCV condition are plotted against the corresponding voltages where the EIS data recorded [see Fig.~\ref{EIS}(f)]. Interestingly, the diffusion coefficient reaches its peak around the Ti$^{3+}$/Ti$^{4+}$ redox potential, suggesting faster sodium-ion mobility in this region. It gradually decreases in the Mn$^{2+}$/Mn$^{3+}$ range, but remains reasonably high. However, a noticeable drop occurs in the Mn$^{3+}$/Mn$^{4+}$ region, indicating slower diffusion kinetics. This overall trend mirrors what is observed in the GITT analysis above. Notably, the lowest $D_{\text{Na}^+}$ values are seen at the cut-off voltages, likely due to the complete (de)sodiation of the active material, which reduces the available pathways for ion transport. 

Finally, it is to emphasize that the DRT method offers an alternative approach for  analyzing and interpreting the EIS data using a continuous distribution of RC elements across the relaxation time domain. A key advantage of the DRT method is that it does not require a predefined equivalent circuit model, making it free from assuming the initial fitting model. This allows for the separation of distinct charge storage processes based on their specific time constants \cite{Wan_EA_15, Plank_JPS_24}. The calculated DRT patterns are presented in Figs.~\ref{DRT}(a, b) during charge and discharge, respectively, which show the visible change in the resistive components at different states of charge. The resulting DRT function displays peaks that correspond to the resistances associated with individual sub-steps of complex electrochemical processes, with each peak representing a distinct physicochemical process, while the area beneath the peak indicates the resistance associated with that specific process \cite{Plank_JPS_24}. In order to probe the dynamic changes in the resistance of each process, the peak fitting of the DRT curves is done, and the deconvoluted spectra are depicted in Figs.~\ref{DRT}(c1--c7) during charge and Figs.~\ref{DRT}(b1--b6) during discharge. The obtained peaks are labeled from P1--P8 in increasing order of time scale, as shown in Fig.~\ref{DRT}(c7). The P1 and P2 fall in the high-frequency region ($\le$10$^{-4}$s) and consist of ohmic contributions from the different cell components like contact, electrolyte, and electrode resistances. The peak P3 belongs to the activity of the passivation layer/SEI in the range of mid-frequency region (10$^{-4}$--10$^{-3}$s). The peaks P4--P7 are positioned in the mid to low-frequency region (10$^{-3}$--10$^{1}$s) and are the active region of the charge-transfer and double-layer resistances. In contrast, the peak P8 at low-frequency regime ($\ge$10$^{1}$s) is attributed to the diffusion process of sodium ions inside the bulk electrode \cite{Plank_JPS_24, Pati_JPS_24, Malik_JES_24}. The peak positions remain mostly consistent across the entire voltage range, with only minor variations, as illustrated in Fig.~S6 of \cite{SI}. 

The calculated resistance values for all the peaks are plotted as a function of voltage, peaks P1, P2, and P4--P7 in Fig.~\ref{DRT}(e), and for peaks P3 and P8 in Fig.~\ref{DRT}(f). The P1 and P5 peaks appear mostly independent of the state of charge, while the P2 and P6 peaks show no significant variation, aside from a slight linear increase near the lower cut-off. The peak P3 with an average time constant of $\sim$400~$\mu$s is related to the impedance offered by the passivation layer/SEI and shows similar abrupt changes at the lower cut-off voltage of 1.5 V, as observed in EIS data shown in Fig.~\ref{EIS}(b, c), and specifically at 2.15 V during discharge \cite{Mohsin_JPS_22}. This confirms the validation that the voltage drop is produced by an increase in polarization of the passivation layer/SEI on the sodium counter electrode instead of a change in the potential of the working electrode \cite{Rudola_EC_14, Malik_JES_24}. The P4 peak positioned at $\sim$5~ms could be the charge transfer resistance at the sodium metal anode, which only offers high impedance when the cell voltage is within the broad inactive redox region between the Ti$^{3+}$/Ti$^{4+}$ and Mn$^{2+}$/Mn$^{3+}$, while remaining stable with low impedance at other voltages \cite{Chen_JPS_23}. Similarly, the charge transfer at the cathode side is represented by peak P7 with average time constant of $\sim$2.5 s, where the impedance reaches its minimum during the Ti$^{3+}$/Ti$^{4+}$ redox process, indicating efficient charge transfer in this voltage region \cite{Mohsin_JPS_22}. The peak P8, with an average time constant of $\sim$50 s, marks the slowest electrochemical process, which is the solid-state diffusion of sodium ions within the bulk electrode, and its trend with the voltage is almost a horizontal mirror image of Fig.~\ref{EIS}(f). The diffusion impedance is lowest during the Ti$^{3+}$/Ti$^{4+}$ redox, increases slightly for Mn$^{2+}$/Mn$^{3+}$, and rises further for the Mn$^{3+}$/Mn$^{4+}$ redox. This trend is consistent with observations from GITT and the Warburg region, reinforcing the reliability of the findings.
As compare to the conventional methods, the DRT analysis provides a clearer picture of how various processes evolve with the state of charge, helping to better understand reaction kinetics, electrode interface activities, performance bottlenecks, and signs of aging \cite{Pati_JPS_24, Mohsin_JPS_22, Chen_JPS_23}.

\section{\noindent~Conclusion}

In summary, our study demonstrates the successful synthesis of Na$_{3.3}$Mn$_{1.2}$Ti$_{0.75}$Mo$_{0.05}$(PO$_4$)$_3$/C cathode material via a sol-gel method, incorporating sodium enrichment and Mo doping to stabilize a high Mn content within the NASICON structure. This dual strategy enhances electrochemical reversibility and enables stable multi-electron redox reactions. The material exhibits excellent cycling stability, delivering 124 mAh/g at 0.1 C and retaining 70\% capacity after 400 cycles at 2 C. The observed sloping voltage profile is attributed to a high density of surface states arising from its nanocrystalline structure, as confirmed by HR-TEM. The carbon-coated nanostructure significantly improves electronic conductivity and ion transport, with diffusion coefficients in the range of $10^{-9}$ to $10^{-11}$~cm$^2$/s, estimated using GITT and CV and a migration energy of $\sim$0.76 eV, predicted from bond-valence site energy mapping. The GITT proved more reliable than CV for evaluating diffusion behavior, effectively isolating pseudocapacitive effects. Additionally, the DRT analysis of EIS data provided a more accurate deconvolution of impedance components across time scales than conventional Nyquist fitting. The consistency among GITT, CV, and DRT analysis confirms the robustness of the analytical approach. These findings validate the effectiveness of the compositional design strategy and underscore the utility of advanced diagnostic techniques in guiding the development of high-performance sodium-ion batteries. 

\section{\noindent~Acknowledgements}
MS thanks CSIR-HRDG for the fellowship support. We acknowledge the DST for financially supporting the research facilities for sodium-ion battery project through {\textquotedblleft}DST-IIT Delhi Energy Storage Platform on Batteries” (project no. DST/TMD/MECSP/2K17/07) and from SERB (now ANRF) through a core research grant (file no.: CRG/2020/003436). We acknowledge IIT Delhi for providing the XPS, FESEM, and HRTEM at the central research facility (CRF), and the XRD and Raman spectroscopy at the department of physics.

\end{document}